\documentclass{amsart}
\usepackage{amssymb}
\usepackage{amsfonts}

\newtheorem{theorem}{Theorem}
\theoremstyle{plain}

\numberwithin{equation}{section}
\input{tcilatex}

\begin{document}
\title[Material elements and substructural interactions]{\textbf{%
Representation of material elements and geometry of substructural
interactions}}
\author{Paolo Maria Mariano}
\address{DICeA, University of Florence, via Santa Marta 3, I-50139 Firenze
(Italy).}
\email{paolo.mariano@unifi.it}
\date{Accepted on 08-06-2007}
\maketitle

\section{Introduction}

This paper collects some characteristic aspects of the general
model-building framework of the mechanics of complex bodies, that are bodies
in which the material substructure influences prominently the gross behavior
through interactions conjugated with substructural changes. Throughout I
review critically some results already published in [4], [10], [11] and yet
in print [12], [13], and add critical remarks. The emphasis is on issues
that are foundational in nature: the geometrical description of the material
elements, their energetic characterization, the representation of
interactions, conditions for the existence of ground states in conservative
setting.

In its primitive meaning, a body can be regarded as an abstract set $%
\mathfrak{B}$ collecting \emph{material elements}, each one being the
smallest piece of matter characterizing the material composing the body. The
basic issue is the `representation' of such a set, a representation obtained
by mapping $\mathfrak{B}$\ in some other set, attributing in this way
(geometrical) structure to $\mathfrak{B}$\ itself.

In the standard format of continuum mechanics (see the treatises [19],
[20]), the geometrical representation adopted is the minimal one: each $%
\mathfrak{e}\in \mathfrak{B}$ is individuated only by a place $x$ by means
of a bijective map $k_{p}$ from $\mathfrak{B}$ into the three-dimensional
space $\mathcal{E}^{3}$ (or $\mathbb{R}^{3}$, that is perhaps more
convenient for some developments below), with the assumption that the whole $%
\mathfrak{B}$ is mapped onto a regular region $\mathcal{B}$ which is called 
\emph{apparent shape} (\emph{place }or \emph{configuration} for short).
Regularity is intended here in the sense that $\mathcal{B}$ is assumed to be
a bounded domain with boundary $\partial \mathcal{B}$ of finite
two-dimensional measure, a boundary where the outward unit normal $n$ is
well defined to within a finite number of corners and edges. In this way one
considers the generic material element as a windowless box, a \emph{monad}
in the words of Leibniz. No information on the shape of the material
element, the shape of its internal structure, is accounted for.

However, evidences of condensed matter physics indicate that the material
elements are groups of entangled molecules, simple or complex pieces of
crystalline structures, stick molecules dispersed in a ground fluid etc.,
depending on physical circumstances envisaged. In all these cases, when
substructural changes determine non-negligible interactions, the standard
representation of bodies is too minimalist. The material element is in
essence a \emph{system} rather than a windowless box. In representing $%
\mathfrak{B}$, then, a map attributing to each material element a \emph{%
morphological descriptor} of its (inner) substructure has to be defined.

Various choices of the morphological descriptors can be made: Elements of
the projective plane may indicate locally the prevailing direction of
nematic order in liquid crystals [5]; scalars may be added to account for
the degrees of orientation [6], prolation and/or triaxiality [3]. Vectors of
the unit sphere $S^{2}$ may represent magnetic spins [1]. Second order
tensors are measures of independent deformations of macromolecules embedded
in a melt [14]. Three-dimensional vectors account for local (random in
essence) atomic rearrangements in quasi-periodic alloys [11]. The taxonomy
of special models is vast, so is the set of results available in single
special cases.

However, in considering such a taxonomy, one becomes aware that to construct
the essential structures of these models, at least at the level of first
principles such as primitive (weak or strong) balances of actions, it is not
necessary to render precise the nature of the morphological descriptor
except assuming that it is an element of a finite-dimensional differentiable
manifold $\mathcal{M}$. Consequently, in representing $\mathfrak{B}$, in
addition to the map $k_{p}$, another map $k_{m}:\mathfrak{B}\rightarrow 
\mathcal{M}$ assigns to each material element the morphological descriptor
(often called order parameter) of its substructural shape. Such a point of
view has been introduced by G. Capriz in the late 1980's [2].

In general I say that the mechanics associated with maps between manifolds
is a unifying setting for existing models of special classes of complex
bodies and, above all, a model-building framework for describing the
mechanical behavior of exotic material which are sometime the results of new
industrial manufactures.

\section{Transplacements and morphological descriptor maps}

Once a reference place $\mathcal{B}_{0}:=k_{p,0}\left( \mathfrak{B}\right) $
is selected (for convenience) in $\mathbb{R}^{3}$, any other actual place $%
\mathcal{B}$ is considered to be achieved in an isomorphic copy\ $\mathbb{%
\hat{R}}^{3}$ of $\mathbb{R}^{3}$ itself by means of a \emph{transplacement}
map $u:=\hat{k}_{p}\circ k_{p,0}^{-1}$, where $\hat{k}_{p}:=i\circ k_{p}$, $%
i $ the isomorphism between $\mathbb{R}^{3}$ and $\mathbb{\hat{R}}^{3}$, so
that below $\mathcal{B}:=\hat{k}_{p}\left( \mathfrak{B}\right) $. The choice
of $\mathbb{\hat{R}}^{3}$ is motivated by the sake of convenience for the
developments below. The map $\mathcal{B}_{0}\ni x\mapsto u\left( x\right)
\in \mathcal{B}$ is assumed as usual to be \emph{one-to-one} and \emph{%
orientation preserving}. Its derivative $Du\left( x\right) \in Hom\left(
T_{x}\mathcal{B}_{0},T_{u\left( x\right) }\mathcal{B}\right) $ is commonly
indicated by $F$ and is the gradient of deformation.

The inner structure of the material elements is described (at least at a
coarse grained level) by a \emph{morphological descriptor map} $\nu
=k_{m}\circ k_{p,0}^{-1}$ with $\mathcal{B}_{0}\ni x\longmapsto \nu \left(
x\right) \in \mathcal{M}$ assumed differentiable with spatial derivative
indicated by $N:=D\nu \left( x\right) \in Hom\left( T_{x}\mathcal{B}%
_{0},T_{\nu \left( x\right) }\mathcal{M}\right) $.

The map $\left( u,\nu \right) :\mathcal{B}_{0}\mapsto \mathbb{\hat{R}}%
^{3}\times \mathcal{M}$ then describes the gross deformation and the
material substructural morphology. It is convenient to maintain $\mathcal{M}$
as abstract as possible: geometrical structures over it have often a precise
physical meaning so that they have to be attributed to $\mathcal{M}$
carefully under the suggestion of specific physical circumstances under
scrutiny.

Here I do not consider motions for the sake of brevity, then the natural
ambient space for describing the (static) equilibrium behavior of complex
bodies is a fiber bundle $\left( \pi ,\mathcal{B}_{0},\mathcal{Y}\right) $
with $\pi $\ the canonical projection. I consider maps $\eta :\mathcal{B}%
_{0}\longrightarrow \mathcal{Y}$ with $\eta \left( x\right) :=\left(
x,u\left( x\right) ,\nu \left( x\right) \right) $. The first prolongation $%
j^{1}\left( \eta \right) $ of $\eta $ is given by%
\begin{equation}
j^{1}\left( \eta \right) \left( x\right) =\left( x,u\left( x\right) ,F\left(
x\right) ,\nu \left( x\right) ,N\left( x\right) \right) =\left( x,y,F,\nu
,N\right)
\end{equation}%
and is an element of the first jet bundle $J^{1}\mathcal{Y}$ over $\mathcal{Y%
}$.

The choice of not considering motions leaves out the discussion about the
nature of the (independent) kinetic energy that can be attributed (when
appropriate) to the material substructures, a discussion in which two
aspects play an essential role: (\emph{i}) the attribution of a metric
structure to $\mathcal{M}$ and (\emph{ii}) the essential \emph{axiom} that
the overall kinetic energy admits an additive decomposition into the
macroscopic and substructural parts, the latter an axiom used commonly
without underlying its essential (foundational) nature.

\section{Standard and substructural actions}

Distinct parts of a body interact with each other and with the rest of the
environment. Actions are naturally defined as objects power conjugated with
the rate of change of places in the case of Cauchy bodies [19], [20]. When
material complexity accrues, actions power conjugated with substructural
changes have to be accounted for [15].

However, since motions are not considered here, it is necessary to make use
of \emph{virtual rates} selected as appropriate vector fields $h\in C\left( 
\mathcal{B}_{0},\mathbb{\hat{R}}^{3}\right) $ and $\upsilon \in C\left( 
\mathcal{B}_{0},T\mathcal{M}\right) $. I consider the pair $\dot{\tau}%
:=\left( h,\upsilon \right) $ as the rate generated by a semigroup over a
space $\mathfrak{G}$ with elements $\tau :=\left( u,\nu \right) $, a space
specified later.\ Let also $\mathfrak{P}$ be the algebra of \emph{parts} of $%
\mathcal{B}_{0}$, each representative of it is indicated by $\mathfrak{b}$.

For any pair $\tau :=\left( u,\nu \right) $, the generic \emph{power} along $%
\left( u,\nu \right) $\ is such a map $\mathcal{P}:\mathfrak{P}\left( 
\mathcal{B}_{0}\right) \times T\mathfrak{G}\rightarrow \mathbb{R}^{+}$ that $%
\mathcal{P}\left( \cdot ,\tau ,\dot{\tau}\right) $ is additive on disjoint
parts and $\mathcal{P}\left( \mathfrak{b},\tau ,\cdot \right) $ is linear.

Two key points have to be discussed now: (\emph{i}) the explicit
representation of $\mathcal{P}$ and (\emph{ii}) its invariance properties
with respect to changes in observers. As regards the latter, since any
observer is a representation of \emph{all} geometrical environments
necessary to describe the morphology of a body and its motion (or
`sequential' deformations, when motions are not accounted for), the
representation of the manifold of substructural shapes $\mathcal{M}$ must be
involved. For \emph{isometric changes in observers} one has to consider the
infinitesimal generators of the action of $\mathbb{\hat{R}}^{3}\ltimes
SO\left( 3\right) $ over the ambient space $\mathbb{\hat{R}}^{3}$ and of the
same copy of $SO\left( 3\right) $ over $\mathcal{M}$, by defining then $%
h^{\ast }:=h+c+q\times x$, with $c\in \mathbb{\hat{R}}^{3}$ and $q\times \in
so\left( 3\right) $, and $\upsilon ^{\ast }:=\upsilon +\mathcal{A}q$, with $%
\mathcal{A}\left( \nu \right) \in Hom\left( \mathbb{\hat{R}}^{3},T_{\nu }%
\mathcal{M}\right) $ so that $\mathcal{A}^{\ast }\left( \nu \right) \in
Hom\left( T_{\nu }^{\ast }\mathcal{M},\mathbb{\hat{R}}^{3}\right) $. For the
sake of brevity I call these changes in observers \emph{semi-classical} and
use the world `semi' to remind that the representation of $\mathcal{M}$ is
involved.

The explicit representation of the power reflects the way in which one
imagines that material elements may interact with each other. If one extends
the standard point of view and assumes the existence of bulk and contact
actions at macroscopic and substructural levels, a rather natural explicit
representation of the \emph{external power} $\mathcal{P}_{\mathfrak{b}%
}^{ext}\left( h\mathbf{,}\upsilon \right) $\ on a generic part $\mathfrak{b}$%
, a power measured over $\left( h\mathbf{,}\upsilon \right) $\ along $\left(
u,\nu \right) $, is given by

\begin{equation}
\mathcal{P}_{\mathfrak{b}}^{ext}\left( h\mathbf{,}\upsilon \right) :=\int_{%
\mathfrak{b}}\left( b\cdot h+\beta \cdot \upsilon \right) \text{ }%
dx+\int_{\partial \mathfrak{b}}\left( Pn\cdot h+\mathcal{S}n\cdot \upsilon
\right) \text{ }d\mathcal{H}^{2}\mathfrak{,}  \label{Pow}
\end{equation}%
with $d\mathcal{H}^{2}$\ the two-dimensional measure, $n$ the normal to $%
\partial \mathfrak{b}$ in all places in which it is defined. At each $x$ the
quantities $b\ $and $\beta $\ are elements of $T_{u\left( x\right) }^{\ast }%
\mathcal{B}\simeq \mathbb{\hat{R}}^{3}$ and $T_{\nu \left( x\right) }^{\ast }%
\mathcal{M}$ respectively, and represent standard and substructural bulk
actions (they split additively in inertial and non-inertial components when
a dynamic setting is accounted for). Standard and substructural contact
actions are represented by the first Piola-Kirchhoff stress $P$ and the
microstress $\mathcal{S}$, which are, at each $x$, elements of $Hom\left(
T_{x}^{\ast }\mathcal{B}_{0},T_{u\left( x\right) }^{\ast }\mathcal{B}\right)
\simeq \mathbb{R}^{3}\otimes \mathbb{\hat{R}}^{3}$ and $Hom\left(
T_{x}^{\ast }\mathcal{B}_{0},T_{\nu \left( x\right) }^{\ast }\mathcal{M}%
\right) \simeq \mathbb{R}^{3}\otimes T_{\nu \left( x\right) }^{\ast }%
\mathcal{M}$, respectively.

\textbf{Axiom}. \emph{At equilibrium the power of external actions is
invariant under semi-classical changes in observers, namely}%
\begin{equation}
\mathcal{P}_{\mathfrak{b}}^{ext}\left( h,\upsilon \right) =\mathcal{P}_{%
\mathfrak{b}}^{ext}\left( h^{\ast },\upsilon ^{\ast }\right)
\end{equation}%
\emph{for any choice of} $\mathfrak{b}$, $c$ \emph{and} $q$.

An immediate theorem follows (see also discussions in Section 8 of [11] and
in references quoted therein). Below \textsf{e} indicates Ricci's
permutation index.

\begin{theorem}
(i) If for any $\mathfrak{b}$\ the vector fields $x\mapsto Pn$ and $x\mapsto 
\mathcal{A}^{\ast }\mathcal{S}n$ are defined over $\partial \mathfrak{b}$
and are integrable there, the integral balances of actions on $\mathfrak{b}$%
\ hold:%
\begin{equation}
\int_{\mathfrak{b}}b\text{ }dx+\int_{\partial \mathfrak{b}}Pn\text{ }d%
\mathcal{H}^{2}=0,  \label{F}
\end{equation}%
\begin{equation}
\int_{\mathfrak{b}}\left( \left( x-x_{0}\right) \times b+\mathcal{A}^{\ast
}\beta \right) \text{ }dx+\int_{\partial \mathfrak{b}}\left( \left(
x-x_{0}\right) \times Pn+\mathcal{A}^{\ast }\mathcal{S}n\right) \text{ }d%
\mathcal{H}^{2}=0.  \label{T}
\end{equation}%
(ii) Moreover, if the tensor fields $x\mapsto P$ and $x\mapsto \mathcal{S}$
are of class $C^{1}\left( \mathcal{B}_{0}\right) \cap C^{0}\left( \mathcal{%
\bar{B}}_{0}\right) $ then%
\begin{equation}
DivP+b=0  \label{Cau1}
\end{equation}%
and there exist a covector field $x\mapsto z\in T_{\nu \left( x\right) }%
\mathcal{M}$ such that%
\begin{equation}
skw\left( PF^{\ast }\right) =\mathsf{e}\left( \mathcal{A}^{\ast }z+\left( D%
\mathcal{A}^{\ast }\right) \mathcal{S}\right)  \label{Skw}
\end{equation}%
and%
\begin{equation}
Div\mathcal{S}-z+\beta =0,  \label{Cap2}
\end{equation}%
with $z=z_{1}+z_{2}$, $z_{2}\in Ker\mathcal{A}^{\ast }$.
\end{theorem}

\begin{itemize}
\item The covector $z$ appearing above is a substructural \emph{self-action}
within the generic material element. In this sense the field $x\mapsto \nu
\left( x\right) $ is self-interacting. If one postulates absence of contact
interactions of substructural nature, by considering in this way the
external power as defined by%
\begin{equation}
\mathcal{P}_{\mathfrak{b}}^{ext}\left( h\mathbf{,}\upsilon \right) :=\int_{%
\mathfrak{b}}\left( b\cdot h+\beta \cdot \upsilon \right) \text{ }%
dx+\int_{\partial \mathfrak{b}}Pn\cdot h\text{ }d\mathcal{H}^{2}\mathfrak{,}
\end{equation}%
a theorem analogous to the previous one holds but the last part reads only
as follows: There exists a covector field $x\mapsto z\in T_{\nu \left(
x\right) }\mathcal{M}$ such that $skw\left( PF^{\ast }\right) =\mathsf{e}%
\mathcal{A}^{\ast }z$ and%
\begin{equation}
\beta =z,  \label{RedCap}
\end{equation}%
with $z=z_{1}+z_{2}$, $z_{2}\in Ker\mathcal{A}^{\ast }$. This last equations
is less trivial than appearing. In fact, when in non-conservative setting
one assumes that $z$ admits an additive decomposition into conservative and
dissipative parts (the latter being linear in the rate of the morphological
descriptor), then the scheme suggested by (\ref{RedCap}) becomes formally
the one of internal variables (see e.g. [18]), differences resting in the
circumstance that internal variables are not observable quantities that do
not describe internal morphologies rather the removal from thermodynamical
equilibrium. Of course, the two points of view can merge one into the other
in appropriate special cases. Take note that, in this case, when $z\in Ker%
\mathcal{A}^{\ast }$ Cauchy stress $\left( \det F\right) ^{-1}PF^{\ast }$ is
symmetric (the condition is sufficient).

\item From the theorem above it appears that a crude integral balance of
substructural actions, namely the integral version of (\ref{Cap2}), has no
geometrical meaning unless $\mathcal{M}$ is embedded in some linear space;
contrary, in fact, the integrand would take values in $T^{\ast }\mathcal{M}$
which is a non-linear space. Moreover, even when $\mathcal{M}$ is embedded,
the integral version of (\ref{Cap2}) does not correspond to any Killing
field of the metric in space. Really the significant balance is the \emph{%
weak balance} of actions%
\begin{equation}
\mathcal{P}_{\mathfrak{b}}^{ext}\left( c+q\times x,\mathcal{A}q\right) =0,%
\text{ \ \ }\forall \mathfrak{b}\in \mathfrak{P},
\end{equation}%
a balance accruing directly from the axiom of invariance of the power. The
power can be in general written in terms of forms over an appropriate space.
A systematic program about the expression of actions in terms of forms has
been initiated and developed by R. Segev (see, e.g., [15], [16], [17], and
references therein).
\end{itemize}

\section{Energy and the existence of ground states}

Once the morphology of the generic material element has been represented
together with the list of potential interactions it may have with the
neighboring fellows and the remaining environment, the local energetic
scenario must be specified: it links morphology and representation of
interactions. In fact, by the standard use of Clausius-Duhem inequality, one
realizes that, at thermodynamical equilibrium, standard and substructural
interactions within the body are determined by derivatives of the energy
with respect to $F$, $\nu $, $N$ (see [2]), under the assumption that $e$ be
differentiable.

Three cases can be discussed.

\begin{enumerate}
\item The generic material element $\mathfrak{e}$ is a closed system with
respect to its substructure in the sense that (\emph{i}) there is no
migration of substructures leaving $\mathfrak{e}$, (\emph{ii}) the material
substructure of $\mathfrak{e}$ does not interact energetically with the
neighboring fellows.

\item The substructure of the generic material element is in energetic
contact with the substructures of the neighboring elements. No migration
occur.

\item The material element is an open system: both energetic contact and
migration of substructures are possible.
\end{enumerate}

Of course the classification above is referred to substructural events. At a
gross scale, in all cases there are interactions between neighboring
material elements considered as a whole, interactions represented by means
of standard tensions.

The attention here is primarily focused on item 2 and, in particular, on the
case in which only conservative phenomena are involved. They are governed by
a 3-form (elastic) energy 
\begin{equation}
\overset{\frown }{\mathfrak{e}}:J^{1}\mathcal{Y\rightarrow \wedge }%
^{3}\left( \mathcal{B}_{0}\right) 
\end{equation}%
of the type 
\begin{equation}
\overset{\frown }{\mathfrak{e}}=e\text{ }dx\mathbf{,}
\end{equation}%
with $e$ a sufficiently smooth density the dependence of which on state
variables is assumed to be given (in isothermal conditions) by%
\begin{equation}
e:=e\left( x,u,F,\nu ,N\right)   \label{en}
\end{equation}%
so that the global energy $\mathcal{E}\left( u,\nu \right) $ of $\mathcal{B}%
_{0}$ is simply%
\begin{eqnarray}
\mathcal{E}\left( u,\nu \right)  &:&=\int_{\mathcal{B}_{0}}\overset{\frown }{%
\mathfrak{e}}\left( j^{1}\left( \eta \right) \left( x\right) \right)   \notag
\\
&=&\int_{\mathcal{B}_{0}}e\left( x,u\left( x\right) ,F\left( x\right) ,\nu
\left( x\right) ,N\left( x\right) \right) \text{ }dx.  \label{Gen}
\end{eqnarray}%
A pair $\left( u,\nu \right) $ satisfying the variational principle%
\begin{equation}
\min_{u,\nu }\mathcal{E}\left( u,\nu \right) \   \label{Var}
\end{equation}%
is called \emph{ground state}. In trying to find minimizers of $\mathcal{E}$%
, constitutive assumptions have to be added: (\emph{i}) the specification of
the functional classes in which one places $u$ and $\nu $, (\emph{ii}) the
`structural' properties of $e$.

\begin{itemize}
\item In the case in which the material element is a closed system with
respect to its substructure, namely when we are within the setting of item 1
of the list above, $e$ is given by $e:=e\left( x,u,F,\nu \right) $ so that
the energetic contribution of the substructure is purely local. Only the
self-action $z$ is present and balances the external bulk action $\beta $ on
the substructure.

\item When the generic material element is an open system (item 3 of the
list above), one has to consider the substructure as a population of
distinct individuals, let say a group of distinct polymeric molecules. Then
it is necessary to add another `morphological' information about the world
inside the material element, namely the \emph{numerosity }of the
substructures, a scalar quantity that satisfies a continuity equation.
Moreover, the substructural migration is intrinsically dissipative: it
generates a loss of information about the local substructural arrangements,
so an increment of configurational entropy and the flux of it is (roughly)
proportional by the chemical potential to the flux of the substructures. The
chemical potential then increases the list of constitutive entries in (\ref%
{en}) together with its gradient. The general treatment of this case is
presented in [10]: a generalized form of Cahn-Hilliard equation arises and
involves a scalar product in the cotangent space of $\mathcal{M}$.
\end{itemize}

The energy $e$ admits commonly an additive splitting of the form $%
e^{i}\left( x,F\mathbf{,}\nu \mathbf{,}N\right) +e^{e}\left( u,\nu \right) $
where $e^{i}\left( x,F,\nu ,N\right) $ is the internal `stored' energy while 
$e^{e}\left( u,\nu \right) $ is the energy of bulk actions. $e^{e}\left(
u,\nu \right) $ splits also in the sum $e_{1}^{e}\left( u\right)
+e_{2}^{e}\left( \nu \right) $ where $e_{1}^{e}\left( u\right) $ is the
potential of standard bulk (gravitational) forces and $e_{2}^{e}\left( \nu
\right) $ the potential of direct bulk actions over the substructure such as
electric fields.

The existence of minimizers for $\mathcal{E}\left( u,\nu \right) $ has been
discussed by G. Modica and myself in [13]. I review here the essential
ingredients of the existence theorem and the theorem itself (or better the
main variant of it).

Constitutive assumptions on the functional nature of the fields involved are
necessary. Preliminarily, it is helpful to remind that if $u:\mathcal{B}%
_{0}\rightarrow \mathbb{\hat{R}}^{3}$ is a Sobolev map, that is an element
of $W^{1,1}\left( \mathcal{B}_{0},\mathbb{\hat{R}}^{3}\right) $, then $%
M\left( Du\right) $ indicates the $3-$vector collecting the minors of $Du$,
i.e. an element of $\Lambda _{3}\left( \mathcal{B}_{0}\times \mathbb{\hat{R}}%
^{3}\right) $. The $n-$current integration $G_{u}$ over the graph of $u$ is
the linear functional on smooth $3-$forms $\omega $ with compact support in $%
\mathcal{B}_{0}\times \mathbb{\hat{R}}^{3}$ defined by%
\begin{equation}
G_{u}:=\int_{\mathcal{B}_{0}}\left\langle \omega \left( x,u\left( x\right)
\right) ,M\left( Du\left( x\right) \right) \right\rangle \text{ }dx\mathbf{,}
\label{CurrSob}
\end{equation}%
so that $\partial G_{u}\left( \omega \right) :=G_{u}\left( d\omega \right) $%
,\ $\omega \in \mathcal{D}^{2}\left( \mathcal{B}_{0}\times \mathbb{\hat{R}}%
^{3}\right) \ $(see [8]).

The deformation $u$ is assumed to be a \emph{weak diffeomorphism} (it is
written $u\in dif^{1,1}\left( \mathcal{B}_{0},\mathbb{\hat{R}}^{3}\right) $%
), in the sense that $u$ is considered a $W^{1,1}\left( \mathcal{B}_{0},%
\mathbb{\hat{R}}^{3}\right) $ map such that (\emph{i}) $\left\vert M\left(
Du\right) \right\vert \in L^{1}\left( \mathcal{B}_{0}\right) $, (\emph{ii}) $%
\partial G_{u}=0$ on $\mathcal{D}^{2}\left( \mathcal{B}_{0}\times \mathbb{%
\hat{R}}^{3}\right) $, (\emph{iii}) $\det Du\left( x\right) >0$ for almost
every $x\in \mathcal{B}_{0}$, (\emph{iv}) for any $f\in C_{c}^{\infty
}\left( \mathcal{B}_{0}\times \mathbb{\hat{R}}^{3}\right) $%
\begin{equation}
\int_{\mathcal{B}_{0}}f\left( x\mathbf{,}u\left( x\right) \right) \det
Du\left( x\right) \text{ }dx\leq \int_{\mathbb{\hat{R}}^{3}}\sup_{x\mathbf{%
\in }\mathcal{B}_{0}}f\left( x,y\right) dy\mathbf{.}
\end{equation}

In particular, the subspace%
\begin{equation}
dif^{r,1}\left( \mathcal{B}_{0},\mathbb{\hat{R}}^{3}\right) :=\left\{ u\in
dif^{1,1}\left( \mathcal{B}_{0},\mathbb{\hat{R}}^{3}\right) |\left\vert
M\left( Du\right) \right\vert \in L^{r}\left( \mathcal{B}_{0}\right)
\right\} ,
\end{equation}%
for some $r>1$, is of special interest below.

As regards the morphological descriptor maps, constitutive assumptions about
the manifold of substructural shapes are first necessary: It is assumed that
(\emph{a}) $\mathcal{M}$ is Riemannian with (at least) $C^{1}-$metric $g_{%
\mathcal{M}}$, and (\emph{b}) covariant derivatives are explicitly
calculated by making use of the natural Levi-Civita connection. A metric
over $\mathcal{M}$ has non-trivial physical meaning with respect to the
representation of the (independent) substructural kinetic energy (when it
exists) and a consequent influence on the representation of the microstress.
Appropriate discussions can be found in [4] and [13]. The connection is
crucial in representing the microstress (take note that such a stress is at
thermodynamic equilibrium the derivative of the energy with respect to $N$%
).\ If no prevalent role is assigned to the Levi-Civita connection, leaving
arbitrary the possibility to select a connection when a specific gauge is
not suggested by the underlying physics, even the parallel transport over
geodetics over $\mathcal{M}$ would result not only in general non-isometric
but even unbounded as a consequence of topological features of $\mathcal{M}$
itself. It is almost trivial to remind that, if a connection would imply an
unbounded parallel transport, the representation of the microstress would
become meaningless. The $C^{1}-$Riemannian structure (assumption (\emph{a})
above) implies that $\mathcal{M}$ can be isometrically embedded in $\mathbb{R%
}^{N}$ by Nash theorem: it is considered here as a \emph{closed submanifold}
in some linear space isomorphic to $\mathbb{R}^{N}$ for some appropriate $N$%
. The use of the Levi-Civita connection implies that the covariant
derivative of $\nu $ is in agreement with the differential of $\nu $ as a
map from $\mathcal{B}_{0}$ into $\mathbb{R}^{N}$. Than the functional
assumption about the map $\nu $ is that it belongs to the Sobolev space $%
W^{1,s}\left( \mathcal{B}_{0},\mathcal{M}\right) $, $s>1$, precisely%
\begin{equation}
W^{1,s}\left( \mathcal{B}_{0},\mathcal{M}\right) :=\left\{ \nu \in
W^{1,s}\left( \mathcal{B}_{0},\mathbb{R}^{N}\right) \text{ }|\text{ }\nu
\left( x\right) \in \mathcal{M}\text{ for a.e. }x\right\} .
\end{equation}

In summary, the minimum problem for the energy introduced above is analyzed
in the functional class%
\begin{equation*}
\mathcal{W}_{r,s}:=\left\{ \left( u,\nu \right) |u\in dif^{r,1}\left( 
\mathcal{B}_{0},\mathbb{\hat{R}}^{3}\right) ,\text{ }\nu \in W^{1,s}\left( 
\mathcal{B}_{0},\mathcal{M}\right) \right\} .
\end{equation*}

Once functional features of the maps $u$ and $\nu $ are (constitutively)
selected, the energy functional $\mathcal{E}$ is extended to $\mathcal{W}%
_{r,s}$ by%
\begin{equation}
\mathcal{E}\left( u,\nu \right) =\int_{\mathcal{B}_{0}}e\left( x\mathbf{,}%
u\left( x\right) ,Du\left( x\right) \mathbf{,}\nu \left( x\right) \mathbf{,}%
D\nu \left( x\right) \right) \text{ }dx\mathbf{,}
\end{equation}%
where $u\left( x\right) $, $Du\left( x\right) $, $\nu \left( x\right) $ and $%
D\nu \left( x\right) $ are the Lebesgue values of $u$, $\nu $ and their weak
derivatives. Assumptions about the structure of the energy $e$, considered
as a map $e:\mathcal{B}_{0}\times \mathbb{\hat{R}}^{3}\times \mathcal{M}%
\times M_{3\times 3}^{+}\times M_{N\times 3}\rightarrow \mathbb{\bar{R}}^{+}$
with values $e\left( x,u,F\mathbf{,}\nu ,N\right) $, are necessary. $%
M_{3\times 3}^{+}\ $and $M_{N\times 3}$ represent the space of $3\times 3$
matrices with positive determinant and the one of $N\times 3$ matrices
respectively.

\begin{itemize}
\item $e$ is assumed to be polyconvex in $F$ and convex in $N$. More
precisely, it is assumed the existence of a Borel function%
\begin{equation}
Pe:\mathcal{B}_{0}\times \mathbb{\hat{R}}^{3}\times \mathcal{M}\times
\Lambda _{3}\left( \mathbb{R}^{3}\times \mathbb{\hat{R}}^{3}\right) \times
M_{N\times 3}\rightarrow \mathbb{\bar{R}}^{+},
\end{equation}%
with values $Pe\left( x,u,\nu ,\xi ,N\right) $, which is (\emph{i}) l. s. c.
in $\left( u,\nu ,\xi ,N\right) $ for a.e. $x\in \mathcal{B}_{0}$, (\emph{ii}%
) convex in $\left( \xi ,N\right) $\ for any $\left( x,u,\nu \right) $, (%
\emph{iii}) and also such that%
\begin{equation}
Pe\left( x,u,\nu ,M\left( F\right) \mathbf{,}N\right) =e\left( x,u,\nu ,F%
\mathbf{,}N\right)
\end{equation}%
for any $\left( x,u,\nu ,F\mathbf{,}N\right) $ with $\det F>0$. In this way
the energy functional becomes%
\begin{equation}
\mathcal{E}\left( u,\nu \right) =\int_{\mathcal{B}_{0}}Pe\left( x\mathbf{,}%
u\left( x\right) ,\nu \left( x\right) \mathbf{,}M\left( F\right) \mathbf{,}%
N\right) \text{ }dx\mathbf{.}  \label{Policon}
\end{equation}

\item It is also assumed that $e$ satisfies the growth condition%
\begin{equation}
e\left( x,u,\nu ,F\mathbf{,}N\right) \geq C_{1}\left( \left\vert M\left(
F\right) \right\vert ^{r}+\left\vert N\right\vert ^{s}\right) +\vartheta
\left( \det F\right)  \label{Growth}
\end{equation}%
for any $\left( x,u,\nu ,F\mathbf{,}N\right) $ with $\det F>0$, $r,s>1$, $%
C_{1}>0$ constants and $\vartheta :\left( 0,+\infty \right) \rightarrow 
\mathbb{R}^{+}$ a convex function such that $\vartheta \left( t\right)
\rightarrow +\infty $ as $t\rightarrow 0^{+}$.
\end{itemize}

Rather detailed remarks about the physical nature of the assumptions above,
assumptions dealing with the influence of the substructure on the local
stability of the material and with the energetic features of the
substructural events, can be found in [13]. With the assumptions above, it
is now possible to analyze the problem of finding minimizers for $\mathcal{E}
$ at least in the case of Dirichlet boundary data. Structure, closure and
consequent compactness results in [8] together with the classical Ioffe's
semicontinuity result allow one to prove the theorem below.

\begin{theorem}
$\left[ 13\right] $ The functional $\mathcal{E}$ achieves the minimum value
in the classes%
\begin{equation}
\mathcal{W}_{r,s}^{d}:=\left\{ \left( u,\nu \right) \in \mathcal{W}%
_{r,s}|u=u_{0}\text{ on }\partial \mathcal{B}_{0,u},\nu =\nu _{0}\text{ on }%
\partial \mathcal{B}_{0,\nu }\right\}
\end{equation}%
and%
\begin{equation}
\mathcal{W}_{r,s}^{c}:=\left\{ \left( u,\nu \right) \in \mathcal{W}_{r,s}%
\text{ }|\text{ }\partial G_{u}=\partial G_{u_{0}}\text{ on }\mathcal{D}%
^{2}\left( \mathbb{R}^{3}\times \mathbb{\hat{R}}^{3}\right) ,\nu =\nu _{0}%
\text{ on }\partial \mathcal{B}_{0,\nu }\right\} .
\end{equation}
\end{theorem}

Above, $\partial \mathcal{B}_{0,u}$ and $\partial \mathcal{B}_{0,\nu }$ are
the portions of the boundary where $u$ and $\nu $ are prescribed; in
particular, the boundary condition $\partial G_{u}=\partial G_{u_{0}}$\ is a
strong anchoring condition (see [8]). On the rest of the boundary, standard
and substructural tractions are assumed to vanish. The physical meaning of
the boundary conditions for the equilibrium problem of complex bodies has
been discussed in detail in [13]. Here I remind only that, although there
are shrewdness that allow one to prescribe in special cases the boundary
value of $\nu $, it appears a hard job to imagine some loading device
prescribing substructural tractions at the boundary. Consequently, the
natural condition could be that the substructural tractions vanish at the
boundary.

The existence theorem above is not accompanied by appropriate regularity
results. Moreover, a Lavrentiev gap phenomenon is not excluded a priori (see
appropriate discussions in [13]; see also [7] and [9]).

Existence results of the type above are also available in terms of
varifolds: in this case minimizers may describe fractured states (the
relevant work is under completion).

\section{Further remarks on standard and substructural actions}

If the minimizer $\left( u,\nu \right) $ is of class $C_{\bar{u}}^{1}\left( 
\mathcal{B}_{0},\mathbb{R}^{3}\right) \times C_{\bar{\nu}}^{1}\left( 
\mathcal{B}_{0},\mathcal{M}\right) $, with $\bar{u}$ and $\bar{\nu}$ the
boundary values of the relevant fields along $\partial \mathcal{B}_{0}$, one
may compute the first variation of $\mathcal{E}\left( u,\nu \right) $ from
the ground state $\left( u,\nu \right) $ by making use of fields $h\in
C_{c}^{1}\left( \mathcal{B}_{0},\mathbb{\hat{R}}^{3}\right) $ and $\upsilon
\in C_{c}^{1}\left( \mathcal{B}_{0},T\mathcal{M}\right) $. Precisely, if one
selects a generic smooth curve $\left( -1,1\right) \ni \varepsilon \mapsto
\nu _{\varepsilon }\in \mathcal{M}$ crossing $\nu $ when $\varepsilon =0$,
then $\upsilon $ is defined by $\upsilon =\frac{d}{d\varepsilon }\nu
_{\varepsilon }\left\vert _{\varepsilon =0}\right. $ and $\upsilon \left(
x\right) \in T_{\nu \left( x\right) }\mathcal{M}$. By exploiting the first
variation $\delta _{h,\upsilon }\mathcal{E}$ of $\mathcal{E}$ from the
ground state $\left( u,\nu \right) $ along the direction $\left( h,\upsilon
\right) $, it is immediate to realize that the map $\varepsilon \mapsto 
\mathcal{E}\left( u+\varepsilon h,\nu _{\varepsilon }\right) $ is
differentiable and the pair $\left( u,\nu \right) $ satisfies the weak form
of Euler-Lagrange equations%
\begin{equation}
\int_{\mathcal{B}_{0}}\left( -b\cdot h+P\cdot Dh+\left( z-\beta \right)
\cdot \upsilon +\mathcal{S}\cdot D\upsilon \right) \text{ }dx=0,
\end{equation}%
for any $\left( h,\upsilon \right) \in C_{c}^{1}\left( \mathcal{B}_{0},%
\mathbb{\hat{R}}^{3}\right) \times C_{c}^{1}\left( \mathcal{B}_{0},T\mathcal{%
M}\right) $, with $\upsilon $ satisfying the condition $\upsilon \left(
x\right) \in T_{\nu \left( x\right) }\mathcal{M}$\ above. Moreover, if $%
\left( u,\nu \right) \in C^{2}\left( \mathcal{B}_{0},\mathbb{R}^{3}\right)
\times C^{2}\left( \mathcal{B}_{0},\mathcal{M}\right) $, then (\ref{Cau1})
and%
\begin{equation}
Div\mathcal{S}-z+\beta =0\text{ \ \ in }T_{\nu }^{\ast }\mathcal{M}
\end{equation}%
correspond to the Euler-Lagrange equations of $\mathcal{E}\left( u,\nu
\right) $ with $P:=\partial _{F}e$, $b:=-\partial _{u}e$, $\mathcal{S}%
:=\partial _{D\nu }e$ and $z-\beta :=\partial _{\nu }e$. In particular, by
exploiting the additive decomposition of $e$ into internal $e^{i}\left(
x,Du,\nu ,D\nu \right) $ and external $e^{e}\left( u,\nu \right) $
components, one gets $z:=\partial _{\nu }e^{i}$ and $\beta :=-\partial _{\nu
}e^{e}$. In this way the specification of the energy eliminates the
indetermination in Theorem 1 due to the presence of the term $z_{2}\in Ker%
\mathcal{A}^{\ast }$. Really, such an indetermination can be eliminated
(under appropriate smoothness conditions) also by using Noether theorem and
requiring covariance, that is invariance of the energy with respect to the
action of the group of automorphisms of the ambient space and the action of
a Lie group over $\mathcal{M}$, precisely the group $Aut\left( \mathcal{M}%
\right) $\ of automorphisms of $\mathcal{M}$. This last requirement can be
considered as invariance with respect to changes in the `representation' of $%
\mathcal{M}$ (see [4] in the conservative case and [11] when substructural
dissipation occurs only within the generic material element). Special
circumstances might require the action over $\mathcal{M}$ of a non-trivial
subgroup of $Aut\left( \mathcal{M}\right) $.

The issue becomes more complicated when one tries to compute the first
variation of $\mathcal{E}$ `around' local minimizers in $\mathcal{W}%
_{r,s}^{d}$. In this case, to avoid problems due to the irregularity of
minimizers, it is convenient (rather than acting directly on the fields) to
make use of horizontal variations induced by maps $\phi \in C_{0}^{1}\left( 
\mathcal{B}_{0},\mathbb{R}^{3}\right) $ which determine, for $\varepsilon $
sufficiently small, diffeomorphisms $\Phi _{\varepsilon }\left( x\right)
:=x+\varepsilon \phi \left( x\right) $ from $\mathcal{B}_{0}$ into itself,
diffeomorphisms that leave unchanged $\partial \mathcal{B}_{0}$. A lower
bound for $Pe\left( x,u,\nu ,M\left( F\right) \mathbf{,}N\right) $ has to be
considered in order to assure coercitivity on $dif^{r,\bar{r}}\left( 
\mathcal{B}_{0},\mathbb{\hat{R}}^{3}\right) \times W^{1,s}\left( \mathcal{B}%
_{0},\mathcal{M}\right) $, $\frac{1}{r}+\frac{1}{\bar{r}}=1$, where 
\begin{equation}
dif^{r,\bar{r}}\left( \mathcal{B}_{0},\mathbb{\hat{R}}^{3}\right) :=\left\{
u\in dif^{r,1}\left( \mathcal{B}_{0},\mathbb{\hat{R}}^{3}\right) \text{ }|%
\text{ }M\left( D\hat{u}\right) \in L^{\bar{r}}\left( \tilde{u}\left( 
\mathcal{B}_{0}\right) \right) \right\} ,
\end{equation}%
and $\tilde{u}$ is the Lusin representative of $u$. The lower bound is
refined with respect to (\ref{Growth}) in the sense that $\vartheta \left(
\det F\right) $ is substituted by $\left\vert M\left( F\right) \right\vert ^{%
\bar{r}}\left( \left( \det F\right) ^{\bar{r}-1}\right) ^{-1}$, i.e. by an
estimate involving the minors of the gradient of the inverse of $u$, that is
the gradient of a $L^{\infty }$ map $\hat{u}$ defined over $\tilde{u}\left( 
\mathcal{\tilde{B}}_{0}\right) $, where $\mathcal{\tilde{B}}_{0}$ is the set
of Lebesgue points of both $u$ and $Du$, and such that both $\hat{u}\circ
u=id_{\mathcal{\tilde{B}}_{0}}$ and $u\circ \hat{u}=id_{\tilde{u}\left( 
\mathcal{\tilde{B}}_{0}\right) }$, and the right and left multiplication of $%
Du$ by $D\hat{u}$ gives rise to the identity. The existence of $\hat{u}$ is
assured by a structure theorem in [8]. Upper bounds are necessary for $Pe$
and its derivatives to assure that the map $\varepsilon \rightarrow \mathcal{%
E}_{\varepsilon }$ be differentiable at zero with derivatives bounded in $%
L^{1}$. By evaluating the effects of horizontal variations, it follows that
(see details in [13]) both $\left( Du\left( x\right) \right) ^{\ast
}\partial _{F}e$ and $\left( D\nu \left( x\right) \right) ^{\ast }\partial
_{N}e$ belong to $L^{1}\left( \mathcal{B}_{0}\right) $, and%
\begin{equation}
Div\mathbb{P-}\partial _{x}e=0  \label{Conf}
\end{equation}%
in distributional sense, with $\mathbb{P}:=eI-F^{\ast }P-N^{\ast }\mathcal{S}%
\in Aut\left( \mathbb{R}^{3}\right) $ the extended Hamilton-Eshelby tensor
valid for complex bodies (see [4] and references therein).

\begin{itemize}
\item Equation (\ref{Conf}) is the bulk balance of configurational forces in
complex bodies in (isothermal) conservative setting. For $C^{2}$ fields, (%
\ref{Conf}) is not essential for analyzing equilibrium problems; contrary,
in presence of evolving bulk defects, a modification of (\ref{Conf})
including an additional driving force furnishes the evolution equation of
the defects themselves. Contrary, the result above (see [8] for a version of
it in the standard non-linear elasticity of simple bodies) points out that (%
\ref{Conf}) is in a sense more essential than expected because, for
irregular minimizers, it furnishes information about the balance of actions
in absence of the Lagrangian representation of the balance of the standard
forces, namely (\ref{Cau1}).

\item Notice that the self force $z$ does not appear in (\ref{Conf})
explicitly. When dissipative substructural effects accrue within the generic
material element, $z$ admits additive decomposition in conservative and
dissipative components. In this case only the dissipative part of $z$,
namely $z^{d}$, appears in the relevant version of (\ref{Conf}) as an
additive term of the type $N^{\ast }z^{d}$ (see [11], [12]).

\item In the case of irregular minimizers, information on substructural
interactions and their balance can be obtained by maintaining $\mathcal{B}%
_{0}$ and $\mathbb{\hat{R}}^{3}$ fixed and altering the manifold of
substructural shapes by means of the action of its group of automorphisms $%
Aut\left( \mathcal{M}\right) $. Precisely, one selects smooth curves $%
\varepsilon \rightarrow \bar{\phi}_{\varepsilon }\in Aut\left( \mathcal{M}%
\right) $, with $\bar{\phi}\in C^{1}\left( \mathcal{M}\right) $, and defines 
$\nu _{\varepsilon }:=\varphi _{\varepsilon }\left( \nu \right) $, by
indicating by $\xi $ the derivative $\frac{d}{d\varepsilon }\nu
_{\varepsilon }\left\vert _{\varepsilon =0}\right. $. Upper and lower bounds
mentioned above assure that [13] (\emph{i}) the map $\varepsilon \rightarrow 
\mathcal{E}\left( u,\nu _{\varepsilon }\right) $ is differentiable at $%
\varepsilon =0$, (\emph{ii}) $\mathcal{S}$ belongs to $L^{1}\left( \mathcal{B%
}_{0},\mathbb{R}^{3\ast }\otimes T^{\ast }\mathcal{M}\right) $ and (\emph{iii%
}) the weak balance%
\begin{equation}
\int_{\mathcal{B}_{0}}\mathcal{S}\left( x\right) \cdot D\xi \left( x\right) 
\text{ }dx+\int_{\mathcal{B}_{0}}\left( z-\beta \right) \left( x\right)
\cdot \xi \left( x\right) \text{ }dx=0,  \label{WeakSubInt}
\end{equation}%
holds for every\emph{\ }$\xi \in C^{0}\left( \mathcal{B}_{0},T\mathcal{M}%
\right) .$

\item In summary, for irregular minimizers the distributional balance of
configurational forces and the weak balance of substructural interactions
are the balances that can be computed in Lagrangian (referential)
representation. The Eulerian (actual) version of the weak balance of
standard forces can be also computed (see [8], [13]).
\end{itemize}

The list of interactions and their possible balances does not end here. When
discontinuity surfaces and line defects occur within a body and are
structured in the sense that they carry own surface and line energy,
respectively, surface and line interactions accrue. Their link with surface
and line energies is discussed in [4], [11], [12] together with their
balances and the invariance properties they satisfy. I remind briefly here
the sole case of a smooth coherent structured discontinuity surface $\Sigma $%
\ which crosses the body and is oriented by a normal vector field $x\mapsto
m:=m\left( x\right) $, $x\in \Sigma $. In referential representation, along
a pair $\left( u,\nu \right) $, a surface standard stress $\mathbb{T}\in
Hom\left( T_{x}^{\ast }\Sigma ,T_{u\left( x\right) }^{\ast }\mathcal{B}%
\right) $, a surface microstress $\mathbb{S}\in Hom\left( T_{x}^{\ast
}\Sigma ,T_{\nu \left( x\right) }^{\ast }\mathcal{M}\right) $ and a surface
self force $\mathfrak{z}\in T_{\nu \left( x\right) }^{\ast }\mathcal{M}$
occur along $\Sigma $, the first two stresses are presumed a priori, the
existence of $\mathfrak{z}$\ can be proven by means of $SO\left( 3\right) $
invariance arguments. Under appropriate smoothness assumptions they satisfy
the surface balances%
\begin{equation}
\left[ P\right] m+Div_{\Sigma }\mathbb{T}=0,
\end{equation}%
\begin{equation}
\left[ \mathcal{S}\right] m+Div_{\Sigma }\mathbb{S}-\mathfrak{z}=0,
\end{equation}%
where $\left[ \cdot \right] $ denotes the jump of the relevant quantity
across $\Sigma $ and $Div_{\Sigma }$ denotes the surface divergence.
Invariance properties of them have been discussed in [4]. The surface
standard and substructural actions listed above concur in the surface
counterpart of (\ref{Conf}) together with the surface energy which is
commonly assumed to be a function of $m$, the surface derivative of $u$, the
surface derivative of $\nu $ and $\nu $ itself, when $\nu $ is continuous
across $\Sigma $. In this equation (which I do not report here) only the
dissipative part of $\mathfrak{z}$ appears, when it exists. The conservative
part of $\mathfrak{z}$ is absent. Relevant proofs can be found in [4] and
[11].

\ \ \ \ \ \ \ \ \ \ \ \ \ \ \ \ \ \ \ \ 

\textbf{Acknowledgements}. This paper is an extended version of a talk I
delivered in August 2006 at Madrid, during the ICM 2006. I wish to thank
Gianfranco Capriz, Giuseppe Modica and Lev Truskinovsky for deep discussions
on the matter we had at Pisa, Firenze and Palaiseau, respectively. My
gratitude goes also (last but not least) to Mirek \v{S}ilhav\'{y} not only
for discussions we had but also for inviting me to contribute this paper.
The support of GNFM-INDAM and MIUR (the latter through the grant 2005085973$-
$"\emph{Resistenza e degrado di interfacce in materiali e strutture}"$-$%
COFIN 2005) is acknowledged. I thank also the "Centro di Ricerca Matematica
Ennio De Giorgi" of the "Scuola Normale Superiore di Pisa" for providing an
adequate environment for scientific interactions.

\section{References}

\begin{description}
\item[{[1]}] \textsc{Brown, W. F. Jr.}, \textit{Micromagnetics}, Wiley, 1963.

\item[{[2]}] \textsc{Capriz, G. }(1989), \emph{Continua with microstructure},
Springer-Verlag, Berlin.

\item[{[3]}] \textsc{Capriz, G.} and \textsc{Biscari, P.} (1994), Special
solutions in a generalized theory of nematics, \textit{Rend. Mat.}, \textbf{%
14}, 291-307.

\item[{[4]}] \textsc{de Fabritiis, C.} and \textsc{Mariano, P. M.} (2005),
Geometry of interactions in complex bodies, \emph{J. Geom. Phys.}, \textbf{54%
}, 301-323.

\item[{[5]}] \textsc{Ericksen, J. L.} (1960), Theory of anisotropic fluids, 
\textit{Trans. Soc. Rheol.}, \textbf{4}, 29-39.

\item[{[6]}] \textsc{Ericksen, J. L.} (1991), Liquid crystals with variable
degree of orientation, \textit{Arch. Rational Mech. Anal.}, \textbf{113},
97-120.

\item[{[7]}] \textsc{Foss, M., Hrusa, W. J.} and \textsc{Mizel, V. J.}
(2003), The Lavrentiev gap phenomenon in nonlinear elasticity, \emph{Arch.
Rational Mech.\ Anal.}, \textbf{167}, 337-365.

\item[{[8]}] \textsc{Giaquinta, M., Modica, G.} and \textsc{Sou}\v{c}\textsc{%
ek, J.} (1998), \emph{Cartesian currents in the calculus of variations},
voll. I and II, Springer-Verlag, Berlin.

\item[{[9]}] \textsc{Hardt, R.} and \textsc{Lin, F. H.} (1986), A remark on $%
H^{1}$ mappings,\emph{\ manuscripta math.}, \textbf{56}, 1-10.

\item[{[10]}] \textsc{Mariano, P. M. }(2005), Migration of substructures in
complex fluids, \emph{J. Phys. A}, \textbf{38}, 6823-6839.

\item[{[11]}] \textsc{Mariano, P. M.} (2006), Mechanics of quasi-periodic
alloys, \emph{J. Nonlinear Sci.}, \textbf{16}, 45-77.

\item[{[12]}] \textsc{Mariano, P. M.} (2006), Cracks in complex bodies:
covariance of tip balances, \emph{J. Nonlinear Sci.}, in print.

\item[{[13]}] \textsc{Mariano, P. M.} and \textsc{Modica, G.} (2006), Ground
states in complex bodies, \emph{ESAIM-COCV}, in print.

\item[{[14]}] \textsc{Mindlin, R. D.} (1964), Micro-structure in linear
elasticity, \textit{Arch. Rational Mech. Anal.}, \textbf{16}, 51-78.

\item[{[15]}] \textsc{Segev, R.} (1994), A geometrical framework for the
statics of materials with microstructure, \emph{Mat. Models Methods Appl.
Sci.}, \textbf{4}, 871-897.

\item[{[16]}] \textsc{Segev, R.} (2000), The geometry of Cauchy fluxes, 
\textit{Arch. Rational Mech. Anal.}, \textbf{3}, 183-198.

\item[{[17]}] \textsc{Segev, R.} (2004), Fluxes and flux-conjugated stresses,
in \textquotedblleft \emph{Advances in multifield theoriesof continua with
substructure}\textquotedblright , G. Capriz and P. M. Mariano Edts., Birk%
\"{a}user, Basel.

\item[{[18]}] \v{S}\textsc{ilhav}\'{y}\textsc{, M.} (1985), Phase transitions
in non-simple bodies, \emph{Arch. Rational Mech. Anal.}, \textbf{88},
135-161.

\item[{[19]}] \v{S}\textsc{ilhav}\'{y}\textsc{, M.} \ (1997), \emph{The
mechanics and thermodynamics of continuous media}, Springer-Verlag, Berlin.

\item[{[20]}] \textsc{Truesdell, C. A.} and \textsc{Noll, W.} (2004), \emph{%
The non-linear field theories of mechanics}, Third edition, Springer-Verlag,
Berlin.
\end{description}

\end{document}